\documentclass[aps,twocolumn,showpacs]{revtex4}

\usepackage{epsfig}

\newcommand{\be}{\begin{equation}}
\newcommand{\ee}{\end{equation}}

\begin{document}

\title{Vacuum-field level shifts in a single trapped ion mediated
by a single distant mirror}

\author{M. A. Wilson, P. Bushev,  J. Eschner,
F. Schmidt-Kaler, C. Becher, and R. Blatt}

\affiliation{Institut f{\"ur} Experimentalphysik, Universit\"{a}t Innsbruck,
A-6020 Innsbruck, Austria}

\author{U. Dorner}

\affiliation{Institut f{\"ur} Theoretische Physik, Universit\"{a}t Innsbruck,
A-6020 Innsbruck, Austria}

\date{\today}

\begin{abstract}
A distant mirror leads to a vacuum-induced level shift in a laser-excited atom.
This effect has been measured with a single mirror 25~cm away from a single,
trapped barium ion. This dispersive action is the counterpart to the mirror's
dissipative effect, which has been shown earlier to effect a change in the
ion's spontaneous decay [J. Eschner et al., Nature \textbf{413}, 495-498
(2001)]. The experimental data are well described by 8-level optical Bloch
equations which are amended to take into account the presence of the mirror
according to the model in [U. Dorner and P. Zoller, Phys. Rev. A \textbf{66},
023816 (2002)]. Observed deviations from simple dispersive behaviour are
attributed to multi-level effects.
\end{abstract}


\maketitle

Classically, it is well understood how the radiative damping of an oscillating
charge leads to the Lorentzian lineshape of emitted radiation \cite{Heitler}.
Early quantum mechanical treatments of the spontaneous decay of an atom,
however, accounted for natural linewidths by invoking the concept of a
zero-point energy or vacuum field \cite{Dirac1927, Weisskopf1930}. The physical
notion of a vacuum field with dynamics independent of the radiating atom was
propounded by Welton \cite{Welton1948}, and that changes in this field could be
observable was noted by Purcell \cite{Purcell1946}. Whilst the vacuum field is
an intuitive and useful concept in explaining many phenomena in
electrodynamics, including such fundamental effects as the Lamb shift and the
Casimir force, it has been recognized that spontaneous decay can be interpreted
as being induced by a combination of vacuum field fluctuations and the atom's
own radiation reaction, the magnitude of the two contributions depending on the
chosen ordering of operators \cite{Milonni}.


An excited-state atom near a mirror, which is the focus of this paper, was
first treated explicitly by Morawitz \cite{Morawitz1969} who showed classically
and quantum mechanically how the presence of the mirror leads to a modified
damping rate and to a change in the resonance frequency of the atom. This
treatment is formally equivalent to earlier work by Lyuboshitz who considered
the scattering of radiation by two dipole centres \cite{Lyuboshitz19671968}.
Arrays of dipoles are used, for example, in the design of directed antennas in
the radio wave range, and similar ideas now form the basis of photonic bandgap
materials. Extensive QED studies were performed by Barton \cite{Barton} who
also considered ground state shifts which are, however, of a fundamentally
different nature to the resonant radiative shifts of this paper
\cite{NoteOnGroundstate}. The modification of the spontaneous decay rate and
the excited-state energy shift induced by a mirror can be attributed in equal
amount to vacuum fields and radiation reaction \cite{Dalibard1982,
Meschede1990}, which leads to a transparent interpretation and an intuitive
reconciliation with the classical theory \cite{Hinds1991, Haroche1992,
Hinds1994}.

Since the first experimental observation of modified spontaneous decay
\cite{Drexhage1974}, the study of changes in the magnitude and spectral
composition of atomic resonance fluorescence due to the presence of nearby
dielectric boundaries has developed to a high level of technical control.
Seminal experiments with atomic beams traversing optical resonators
\cite{Heinzen1987} or Rydberg atoms in microwave cavities \cite{Brune1994} are
now being continued into the domain where single atoms are transiently stored
in high-finesse resonators \cite{Pinkse2000, Hood2000}, and cavity-induced
cooling has been observed \cite{Chan2003}. Unlike these experiments with
2-mirror cavities and non-localized atoms, in this paper we report on the
effect of only a single mirror, which is some 25~cm away, on a single trapped
$^{138}$Ba$^+$ ion which is localized much better than the optical wavelength.

Recently, single ions have been trapped inside optical cavities and shown to
interact deterministically \cite{Guthoehrlein2001, Mundt2002} and coherently
\cite{Mundt2002a} with the cavity field. The action of a single mirror has
previously been shown to lead to position-dependent enhanced and inhibited
spontaneous decay \cite{Eschner2001}. In addition to this dissipative effect,
here we report the dispersive action of the mirror, an energy shift of the
excited atomic level, which is found to be in accordance with a theoretical
treatment of the experimental set-up \cite{Dorner2002}, and which furthermore
shows some peculiar multi-level features.

\begin{figure}[htb]
\epsfig{file=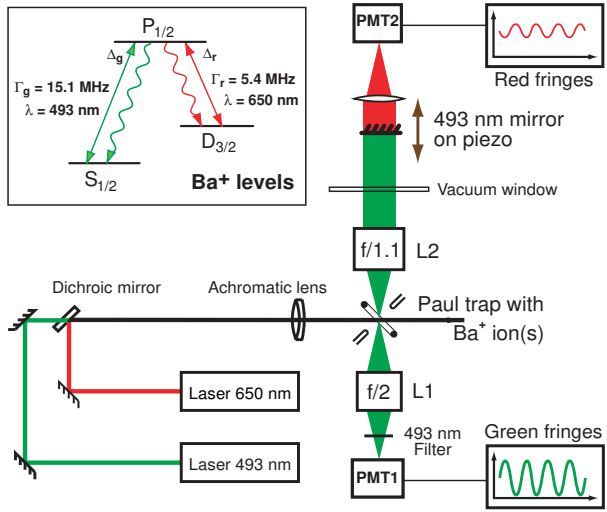, scale=0.55} 
\caption{\label{setup} Schematic of the experiment and relevant energy levels
of Ba$^+$. Two lasers at 493~nm and 650~nm excite a single trapped Ba$^+$ ion
continuously. The photons on the photomultipliers PMT1 and PMT2 are counted in
0.1~s intervals. Further details of the laser systems and the setup are found
in \protect{\cite{Raab1999, Raab2000, Eschner2001}}. }
\end{figure}

The experimental setup and partial level scheme of $^{138}$Ba$^+$ are shown in
Fig.~\ref{setup}. A single $^{138}$Ba$^+$ ion is confined in a miniature Paul
trap with radial and axial secular frequencies of about 1 and 2~MHz,
respectively. Narrowband tunable lasers at 493~nm (green) and 650~nm (red)
drive the $^{2}$S$_{1/2}$ to $^{2}$P$_{1/2}$ and $^{2}$D$_{3/2}$ to
$^{2}$P$_{1/2}$ transitions and Doppler-cool the ion. Laser frequencies are set
close to resonance and laser intensities are around saturation. A macroscope
lens L1 (f-number 2) collects part of the resonance fluorescence, the green
photons of which are detected on PMT1 at count rates around $15 \times 10^{3}$
cps. The resonance fluorescence emitted into the opposite direction is
collimated by lens L2 (f-number 1.1, wavefront abberations $< \lambda/5$). The
green part of this signal is retro-reflected by a piezo-adjustable plane
mirror, while the red part is transmitted and detected on PMT2 at around $25
\times 10^{3}$ cps.

Lens L2, situated inside the vacuum chamber, and the retro-reflecting mirror,
situated about 25~cm away from the ion, are arranged such that they image the
ion onto itself, i.e.\ the returning light is brought to a focus at the
position of the ion. The ion and its mirror image can be observed visually
through L1. Overlapping the mirror image with the real ion leads to
high-contrast interference fringes at PMT1 upon scanning the ion-mirror
distance with a piezo actuator \cite{Eschner2001}. Fine adjustment of the
overlap is critical in obtaining high-contrast interference fringes.
Disappearance of the interference pattern upon tilting the mirror indicates
that the retro-reflected beam is focussed at the position of the ion to within
2~$\mu$m. The green fringe contrast can be as high as 72\%, limited by the
optical set-up (mirror and window flatness, quality of L2), by the thermal
motion of the ion \cite{Eschner2003} and by acoustic noise between the trap and
the external mirror.


Simultaneously with the green signal on PMT1 we record on PMT2 the red
fluorescence transmitted through the mirror. Varying the mirror-ion distance,
which gives rise to the 493~nm fringes, is also seen to modulate the red light
with the same period \cite{Eschner2001}. Note that this modulation is not an
interference at the red wavelength, as this would lead to a different
modulation period. The reason for the red fringes is the back-action of the
mirror on the atom, i.e.\ that the mirror modifies the vacuum field at the
green wavelength and leads to enhancement and inhibition of spontaneous decay
from the P$_{1/2}$ state. As a consequence, the population of the P$_{1/2}$
level is modulated. Since the mirror reflects only the radiation at 493~nm,
only the decay constant on the $^{2}$S$_{1/2}$ to $^{2}$P$_{1/2}$ transition is
modified, its variation being proportional to the green fringes. In contrast,
the detected red light is a measure of the P-state population, thus revealing
the back-action of the mirror on the atom. Then one could naively expect that
enhancement of spontaneous decay at 493~nm leads to increased de-population of
the upper state and a decrease in the rate of detected 650~nm photons, while
inhibited decay at 493~nm increases the 650~nm count rate. However, instead of
such anti-correlation, we observe a phase between the green and the red
modulation which varies with the laser detuning and takes all values between
correlation (phase close to 0 or $2\pi$) and anti-correlation (phase $\pi$).

\begin{figure}[htb]
\epsfig{file=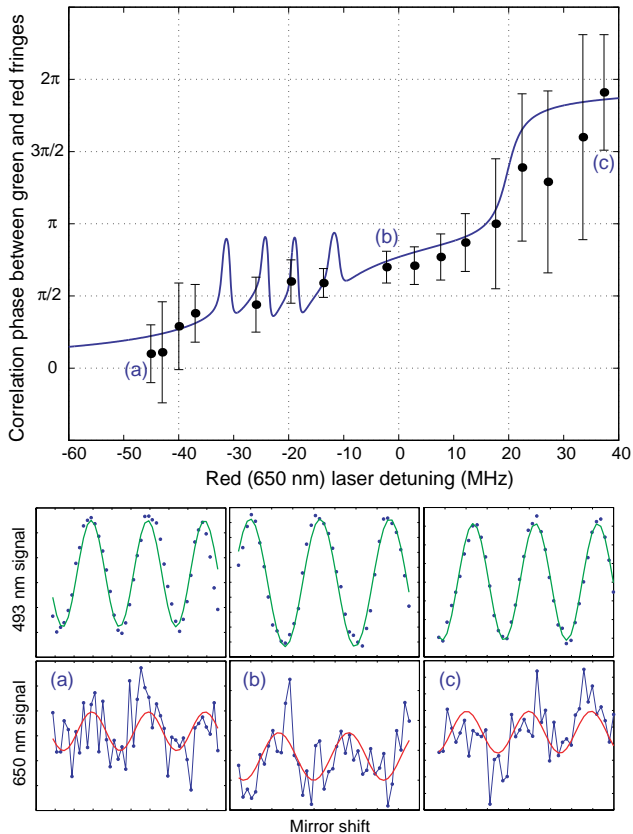, scale=0.16} 
\caption{\label{corrphase} Large graph: Correlation phase between the observed
fringes in the green and in the red fluorescence, vs.\ detuning of the red
laser. The line is a calculation using 8-level Bloch equations. The narrow peak
structures are caused by dark resonances \protect{\cite{Schubert1995,
Raab1999}}. For the three data points marked a), b), c), the smaller graphs
display the simultaneously recorded green (top) and red (bottom) fringes,
showing how the correlation phase varies between 0 and $2\pi$.}
\end{figure}

A plot of this correlation phase as a function of the detuning of the red laser
is shown in Fig.~\ref{corrphase}. For large negative or positive detunings, the
green and red fringes are in phase, while with the red laser close to
resonance, anti-correlation is observed. As we will show now, this dependence
of the correlation phase is a direct consequence and an experimental
verification of the energy shift of the P$_{1/2}$ state which goes along with
its modified decay rate.

We describe the dynamics of the laser-driven barium ion by optical Bloch
equations. In the presence of a magnetic field the degeneracy of the S, P and D
states is lifted so that a realistic model requires all eight Zeeman sublevels
to be taken into account \cite{Schubert1995}. Without the action of the mirror,
the 8-level Bloch equations are routinely used to determine the saturation
parameters of each laser and the detuning of the 493~nm laser, from an
excitation spectrum as the detuning of the 650~nm laser is scanned over
resonance, see Ref.~\cite{Raab1999}. Within this model the decay constants on
the green and red transitions are given by $\Gamma_{g}$ and $\Gamma_{r}$, and
the detunings of the 493 and 650~nm lasers by $\Delta_{g}$ and $\Delta_{r}$,
respectively. To account for the action of the mirror, the model is amended
according to the results of Ref.~\cite{Dorner2002}: the decay constant on the
P$_{1/2}$ to S$_{1/2}$ transition is modified as
\be \Gamma_{g}\rightarrow \Gamma_{g}\left(1-\varepsilon\cos(2kl)\right)~, \ee
and the detunings become
\be \Delta_{g,r}\rightarrow \Delta_{g,r}-(\varepsilon\Gamma_{g}/2)\sin(2kl)~,
\label{detuning} \ee
where $\varepsilon$ is the effective fraction of 4$\pi$ solid angle subtended
by L2 \cite{footnote_on_epsilon}, $k = 2\pi/493$~nm and $l$ is the ion-mirror
distance. It is easily seen that Eq.~(\ref{detuning}) actually represents an
energy shift of the P$_{1/2}$ level. The shift is independent of the laser
intensities, i.e.\ it remains even when the lasers are switched off. Thus it is
not explained by a real AC stark shift due to the back-reflected 493~nm
photons. Instead, its dependence on $\varepsilon$ shows that it is caused by
the mere presence of the mirror and the associated modification of the vacuum
field and the radiation reaction of the atom. Our value of $\varepsilon \approx
2\%$ implies a level shift of approximately $\pm 150$~kHz.

The correlation phase calculated from the amended Bloch equations is plotted in
Fig.~\ref{corrphase}. It describes well the experimental data. In particular,
it must be noted that without the level shift correction of
Eq.~(\ref{detuning}), the calculated dependence would have no other values than
$\pi$ or 0. This can also be seen from an expansion of the excited state
population $\mathcal{P}_{\rm P}$ for small $\varepsilon$, which leads to an
expression of the form
\be \mathcal{P}_{\rm P} = A_1 + \varepsilon \left( A_2\cos(2kl) + A_3\sin(2kl)
\right)~, \ee
where the quantities $A_1,A_2,A_3$ depend on the laser intensities and
detunings \cite{Dorner2002}. Without a level shift, the factor $A_3$ would
vanish and the signals would always be (anti-) correlated.

The data in Fig.~\ref{corrphase} represent 7 hours of near-constant
interrogation of a single barium ion. For each 650~nm laser detuning, up to 80
green and red interference periods were recorded simultaneously while the
ion-mirror distance was varied. The correlation phase was then determined from
the data afterwards. The 493~nm laser is stable to about 1~MHz over long
periods whilst a red laser reference cavity drift rate of $\sim 2$~MHz/hr
limited the accuracy to which the 650~nm laser detuning could be determined.
The data in Fig.~\ref{corrphase} have detunings accurate to within $\pm$1~MHz.



\begin{figure}[htb]
\epsfig{file=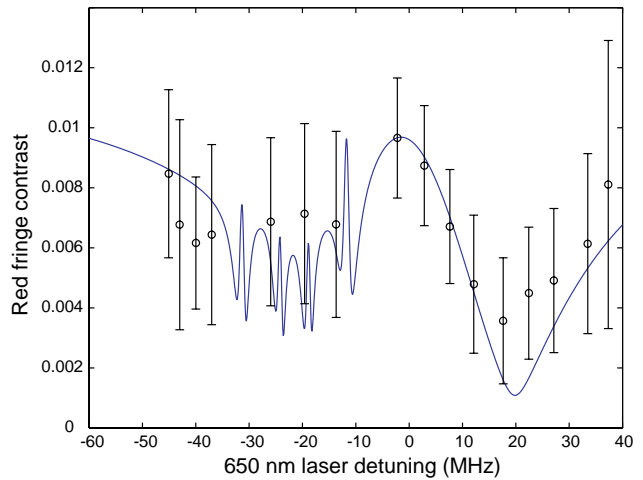, scale=0.45} 
\caption{\label{redcontrast} Red fringe contrast vs.\ detuning of the 650~nm
laser for the data set of Fig.~\ref{corrphase}. The curve is calculated from
Bloch equations with an effective solid angle of $\varepsilon = 1.6\%$ of
$4\pi$. The error bars are due to shot-to-shot variations. The maximum observed
contrast of a single shot corresponds to $\varepsilon = 3\%$.}
\end{figure}

The red fringe contrast also varies with red laser detuning. It reaches values
up to 2.5\%, but for the laser parameters used in the data of
Fig.~\ref{corrphase}, it goes to a minimum for $\Delta_r \approx +20$~MHz.
This, in conjunction with poor Doppler cooling at positive detunings, accounts
for the large measurement errors in the correlation phase in this regime. The
measured variation of red fringe contrast is shown in Fig.~\ref{redcontrast}.
From a comparison with the expected dependence, again calculated with the Bloch
equations, we estimate that optimally the mirror subtended an effective solid
angle of $\varepsilon$=3.2\% of $4\pi$, but the typical value is about 2\%.


\begin{figure}[htb]
\epsfig{file=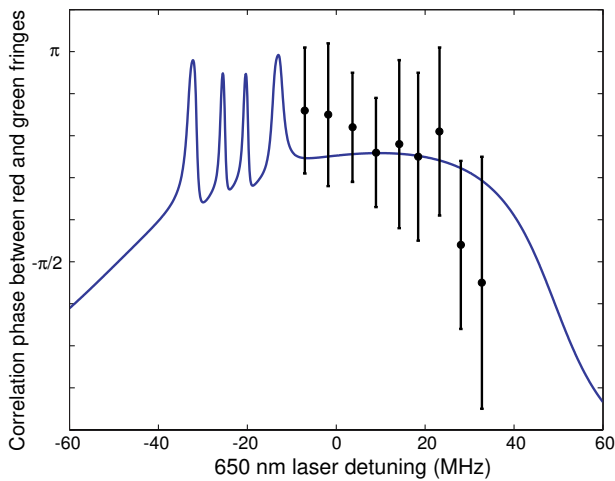, scale=0.45} 
\caption{\label{phasedown} Measurement of correlation phase vs.\ red detuning
in a case when for large positive detunings the correlation phase goes back to
zero rather than reaching $2\pi$. Due to sub-optimal laser cooling, resulting
from high red laser power, the fringe contrast goes down which leads to the
large error bars. }
\end{figure}

While all observations are well described by the model, thus verifying that the
level energies and decay rates are indeed modified by the distant mirror, we
also find peculiar features which are due to the multi-level structure of the
atom and would not appear in a simple two-level system. The red fringe contrast
can fall completely to zero for a specific positive detuning of the 650~nm
laser and a particular ratio of the laser intensities. In other words,
modification of the decay constant on the green transition may have no effect
on the excited-state population. If from that particular set of parameters, the
red laser intensity is increased further, the dispersive dependence of
correlation phase vs.\ detuning shown in Fig.~\ref{corrphase} changes shape. At
large positive $\Delta_r$ it is seen to return to zero instead of going up to
$2\pi$. An example is displayed in Fig.~\ref{phasedown}. While it can be
suspected that an interplay of modified decay and optical pumping is
responsible for this behavior, the detailed underlying causes are the subject
of future study.

In summary, we have experimentally verified that a distant mirror shifts the
energy of the excited atomic levels by modifying the electromagnetic vacuum
around an atom and the atom's radiation reaction. While for non-localized atoms
traversing resonators this effect has been observed earlier, in our case only a
single mirror is used, and the spatial dependence of the energy shift is
exploited by using a single trapped ion whose position is controlled on the
sub-wavelength scale. Due to the multi-level structure of the ion, further
effects arise which distinctly deviate from the behavior of simpler systems.
While our experiment is designed to produce large level shifts, in the range of
few 100~kHz, already much smaller values, which may accidentally appear due to
nearby dielectric objects, would be relevant in precision measurements or
optical clocks with single atoms.


\textbf{Acknowledgements.} This work was supported by the Austrian Science Fund
(FWF, SFB15), by the European Commission (QUEST network, HPRNCT- 2000-00121,
QUBITS network, IST-1999-13021), and by the Institut f\"ur Quanteninformation
GmbH. P.~B.\ acknowledges helpful discussions with P.\ Fedichev.

\end{document}